\documentclass[reprint,nofootinbib,amsmath,amssymb,showkeys,aps,
preprintnumbers]{revtex4-2}

\usepackage{ulem}
\usepackage{graphicx}
\usepackage{epstopdf}
\usepackage{comment}
\usepackage{lipsum}
\usepackage{dcolumn}
\usepackage{bm}
\usepackage{comment}
\usepackage[T1]{fontenc}
\usepackage{mciteplus}
\usepackage{color}
\usepackage{enumitem}
\usepackage{subfigure} 
\usepackage{braket}
\usepackage{enumitem}
\usepackage[colorlinks]{hyperref}
\hypersetup{linkcolor=blue,filecolor=blue,urlcolor=blue,citecolor=blue}
\newcommand{\lsim}{\raisebox{-0.13cm}{~\shortstack{$<$
\\begin{equation}-0.07cm]  $\sim$}}~}
\newcommand{\gsim}{\raisebox{-0.13cm}{~\shortstack{$>$
\\begin{equation}-0.07cm] $\sim$}}~}

\arraycolsep=1.4pt\def\arraystretch{2.2}

\def\d{\delta}

\def\ep{\epsilon}

\def\f{\frac}

\def\l{\left}

\def\m{\mu}
\def\n{\nu}

\def\p{\partial}
\def\r{\right}
\def\s{\sigma}

\def\vp{\varphi}
\def\sech{\rm sech}

\def\be{\begin{equation}}
\def\ee{\end{equation}}
\def\bes{\begin{equation*}}
\def\ees{\end{equation*}}
\def\bea{\begin{eqnarray}}
\def\eea{\end{eqnarray}}
\def\ba{\begin{array}}
\def\ea{\end{array}}
\def\bc{\begin{center}}
\def\ec{\end{center}}
\def\bl{\begin{flushleft}}
\def\el{\end{flushleft}}
\def\br{\begin{flushright}}
\def\er{\end{flushright}}
\def\bi{\begin{itemize}}
\def\ei{\end{itemize}}

\def\bt{\begin{tabular}}
\def\et{\end{tabular}}
\newtheorem{question}{Question}
\def\bq{\begin{question}}
\def\eq{\end{question}}
\newtheorem{definition}{Def}
\def\bd{\begin{definition}}
\def\ed{\end{definition}}
\newtheorem{answer}{Answer}
\def\ban{\begin{answer}}
\def\ean{\end{answer}}
\newtheorem{possibleanswer}{Possible answer}
\def\bpa{\begin{possibleanswer}\normalfont}
\def\epa{\end{possibleanswer}}
\usepackage{mathtools}
\usepackage{comment}
\usepackage{enumerate}
\usepackage{mathrsfs}
\usepackage{units}
\usepackage{graphicx,amsfonts,amssymb,amsthm,amsmath,psfrag}
\usepackage[toc,page]{appendix}
\usepackage{slashed}
\usepackage{mathtools}
\usepackage{tikz}
\usepackage{booktabs}
\usepackage{bm}         
\usepackage{color}
\usepackage{amssymb}
\usepackage{amsmath}     
\usepackage{psfrag}
\usepackage{graphicx}
\usepackage{epstopdf}
\usepackage{dcolumn}
\usepackage{bm}
\usepackage[utf8]{inputenc}
\usepackage{enumitem}
\usepackage{braket}
\usepackage[color=orange]{todonotes}

\definecolor{asparagus}{rgb}{0.53, 0.66, 0.42}
\hypersetup{linkcolor=blue,filecolor=blue,urlcolor=blue,citecolor=blue}
\usepackage{bbding}
\usepackage{amsmath}
\usepackage{multirow}

\usepackage{xspace}
\newcommand*{\ie}{i.e., }

\usepackage{xifthen}
\newcommand*\diff{\mathrm{d}} 
\newcommand*\ldiff[2][]{ \ifthenelse{\isempty{#1}}{ \diff
#2}{\diff^#1#2} \,} 
\let\limitint\int 
\renewcommand{\int}{\limitint \!} 

\begin{document}
 \vspace*{-5mm}
\begin{flushright}

\end{flushright}
\vspace{.3cm}
\thispagestyle{empty}
\title{The geometry of inflationary observables: \\
lifts, flows, equivalence classes}

\author{Georgios K. Karananas$^{1}$}
\email{georgios.karananas@physik.uni-muenchen.de}
\author{Marco Michel$^2$}
	\email{michelma@post.bgu.ac.il}
\author{Javier Rubio$^3$}
\email{javier.rubio@ucm.es}
\affiliation{$^1$ Arnold Sommerfeld Center,
Ludwig-Maximilians-Universit\"at M\"unchen,  Theresienstra{\ss}e 37,
80333 M\"unchen, Germany}
\affiliation{$^2$ Department of Physics, Ben-Gurion University of the
Negev, Beer-Sheva 84105, Israel}
\affiliation{$^3$ Departamento de F\'isica Te\'orica and Instituto de
F\'isica de Part\'iculas y del Cosmos (IPARCOS-UCM), Universidad
Complutense de Madrid, 28040 Madrid, Spain}

\preprint{LMU-ASC~26/23}
\preprint{IPARCOS-UCM-23-053}

\begin{abstract}
The Eisenhart lift allows to formulate the dynamics of a scalar field in
a potential as pure geodesic motion in a curved field-space manifold
involving an additional fictitious vector field. Making use of the
formalism in the context of inflation, we show that the main
inflationary observables can be expressed in terms of the geometrical
properties of a two-dimensional uplifted field-space manifold spanned by
the time derivatives of the scalar and the temporal component of the
vector. This allows to abstract from specific potentials and models and
describe inflation solely in terms of the flow of geometric quantities.
Our findings are illustrated through several inflationary examples
previously considered in the literature.
\end{abstract}

\maketitle

\section{Introduction}

Inflation is nowadays the leading paradigm for explaining, not only the
homogeneity and isotropy of the Universe on large scales, but also the
generation of an almost scale invariant spectrum of primordial density
fluctuations seeding structure formation. The only missing piece of
evidence is a primordial B-mode polarization pattern in the Cosmic
Microwave Background. This is the main objective of current or future
planned experiments such as BICEP3~\cite{Wu:2015oig},
LiteBIRD~\cite{Matsumura:2013aja} and the Simons
Observatory~\cite{SimonsObservatory:2018koc}.

The vast majority of inflationary models makes use of a scalar field,
the inflaton, rolling down (slowly) a sufficiently flat potential.
While the possibilities for implementing this idea are certainly more
than extensive, a wide range of scenarios can be grouped together under
certain equivalence classes involving generically attractor behaviors.
This allows to identify universal properties dictating the dynamics of
fields and observables while ridding of unimportant
model-specific details. Examples of this are the so-called
$\alpha$-attractor settings~\cite{Galante:2014ifa,Artymowski:2016pjz},
the proposal based on the effective equation-of-state
parameter~\cite{Mukhanov:2013tua}, or the $\beta$-function formalism
parametrizing the departure from a nearly scale-invariant regime
\cite{Binetruy:2014zya}. In the same spirit, we propose here a novel,
geometric, point of view that condenses the essential features of single
field inflation in the curvature(s) associated with the target manifold
of the  ``Eisenhart-lifted'' version of the  theory.

In  classical mechanics, the motion of a particle in a given potential
may be described as the geodesic of a free particle moving in a
nontrivial higher dimensional manifold---this is the Eisenhart
lift~\cite{eisenhart1928dynamical}. Both representations are  equivalent
(at least classically) and lead exactly to the same physical
predictions. The generalization of the Eisenhart lift for field theories
(in flat and curved spacetimes) was recently developed in
Ref.~\cite{Finn:2018cfs}, where it was shown that its successful
implementation requires the introduction of a fictitious vector field.
As far as cosmology is concerned, the Eisenhart lift for field theories
has been applied in the context of the initial conditions problem of
inflation, which, in this language, becomes a condition in the kinetic
sector of the uplifted theory~\cite{Finn:2018krt, Finn:2019owf}.

In this short paper, we will make use of this  formalism to classify
observables in an inflationary setting involving a (canonical) scalar
field moving down an arbitrary potential. In particular, by recasting
these scenarios as purely kinetic theories  with an additional
fictitious vector field, we will demonstrate the equivalence of
curvature perturbations at first order in perturbation theory, showing
explicitly that the main inflationary observables  are related to
geometrical objects describing the kinetic manifold of the uplifted
theory.  This opens up a new avenue for model classification and
building,  that we will explicitly illustrate through examples
previously considered in the literature.

This paper is organized as follows. In
Sec.~\ref{sec:eisenhartInflation},  we briefly review the Eisenhart lift
for a scalar field in curved spacetime.  In Sec.~\ref{sec:inflation}, we
discuss the inflationary dynamics, by studying the background evolution
as well as the first order perturbations. The main results of the paper
are presented in Sec.~\ref{sec:observables} and~\ref{sec:lifts}: first we
 show that the inflationary observables can be completely described
in terms of the geometry of the Eisenhart-lifted field-space manifold,
and second we formulate inflation as a fixed-point theory in the space
of field-curvature.  Our conclusions are presented in
Sec.~\ref{sec:conclusion}.

\section{Eisenhart Lift}
\label{sec:eisenhartInflation}

Let us consider the action for a scalar field $\phi$  minimally coupled
to gravity in the mostly plus signature 
\be
\label{eq:minimal_coupling_scalar_act}
S= \int d^4x \sqrt{g} \left[ \frac{M_P^2}{2}R - \frac{1}{2}(\partial
\phi)^2 -V(\phi) \right] \ ,
\ee
with $g=-\text{det}(g_{\m\n})$  the metric determinant and $V(\phi)$ an
arbitrary potential. As shown in Ref.~\cite{Finn:2018cfs}, the dynamics
of this system can be alternatively described by the following action
involving an additional vector field $B_\mu$ with non-canonical kinetic
term, namely~\cite{Finn:2018cfs}
\be
\label{eq:minimal_coupling_equiv_action}
S_{\rm eq}=\f 1 2\int d^4x \sqrt{g} \left [ M_P^2 \, R -  
(\partial \phi)^2 +  \frac{ M^4}{V(\phi)}(\nabla \cdot B)^2
\right] \ , 
\ee
with $M$ an arbitrary constant parameter with the dimension of mass and
$\nabla \cdot B=\nabla_\m B^\m$. The equations of motion following from
this equivalent action take the form
\bea
&&M_P^2\, G_{\mu \nu}  = \partial_\mu \phi \partial_\nu \phi-g_{\mu \nu}
\left[\frac{1}{2} (\partial \phi)^2 + F^2\, V(\phi)\right]\nonumber\\
&&\qquad\qquad\qquad\quad+\sqrt{2} \, M^2 \left[B_{(\m}\p_{\n)}- g_{\mu
\nu} B^\lambda \p_\lambda \right]  F\ ,
\label{eq:einstein_full}\\
&& \Box \, \phi = F^2  V'(\phi)\ ,\label{eq:phifield_full} \\
&& \p_\m  F = 0 \ ,  \label{eq:Bfield_full}
\eea
with $G_{\m\n}=R_{\m\n}-\f 1 2 g_{\m\n}R$ the Einstein tensor, $\square=
g^{\rho\s}\nabla_\rho\nabla_\s$ the covariant  d’Alembertian, primes
standing for differentiation w.r.t. the field,  and 
\be\label{FB}
F= \frac{ M^2\, \nabla \cdot B}{\sqrt{2} \, V(\phi)}\ .
\ee 
Note that Eq.~(\ref{eq:Bfield_full}) is actually a constraint equation
from which $\nabla\cdot B$ is determined; indeed, for $F=1$ we find 
\be
\label{eq:constraintB}
\nabla\cdot B =  \f{\sqrt 2 V(\phi)}{M^2} \ . 
\ee
Plugging this into Eq.~(\ref{eq:phifield_full}), we can easily conclude
that the usual Einstein and Klein-Gordon equations associated with the
starting action~\eqref{eq:minimal_coupling_scalar_act} are indeed
recovered. As we will see below, the $\sqrt{2}$ normalization factor in
Eqs.~(\ref{FB}) and~(\ref{eq:constraintB}) leads to a geodesic motion
parametrized by the coordinate time $t$. A different normalization would
simply lead to a different affine class, with the system evolving slower
or faster. 

\section{Inflation}\label{sec:inflation}
\subsection{Background evolution}
\label{sec:background}

Since we are mainly interested in the cosmological implications of the
Eisenhart formalism, in what follows we will focus on a flat
Friedmann-Lema\^itre-Robertson-Walker (FLRW) metric $\bar
g_{\mu\nu}={\rm diag}(-1,a^2 \delta_{ij})$ with scale factor $a=a(t)$
and $i,j=1,2,3$. Taking the scalar and vector fields to be homogeneous
functions of time at the background level,  
\be
\label{eq:fields_background}
\phi=\bar \phi(t)\,, \hspace{7mm} B_\mu =\bar B_\m(t)  \,,
\ee
the action~(\ref{eq:minimal_coupling_equiv_action})  becomes
\be\label{eq:act2fields}
S_{\rm eq} =\int d^4 x \, a^3\left [\frac{M_P^2}{2}\bar R  +\f 1 2\dot{
\bar{\phi}}^2 +\f 1 2 \frac{M^4}{a^6 \bar V }\dot{ \bar{\chi}}^2 \right
] \ ,
\ee
with 
\be
\bar R = 6   \left(\dot H+2H^2 \right) \ , 
\hspace{7mm} H=\frac{\dot a}{a}\, .
\ee
As usual, the dots denote derivatives w.r.t. the coordinate time,
$\dot{}= \p_t$, and 
\be
\label{eq:def_chi}
\bar V=V(\bar\phi)\,,
\hspace{7mm}
\bar\chi = a^3 \bar B_0\,.
\ee
Let us now write the action~(\ref{eq:act2fields})  as
\be
\label{Smetric}
S_{\rm eq} = \int d^4 x \,\, a^3\left [\frac{M_P^2}{2}\bar R 
+\frac{1}{2}\, M_P^2\,  G_{IJ}\dot\Phi_I \dot \Phi_J \right ] \ ,
\ee
with  $\Phi_I = (\bar\phi,\bar\chi)^T$, and 
\be
G_{IJ} = G_{IJ}(\bar \phi,a)=
\begin{pmatrix}
  1 & 0 \\
  0 & \f {M^4} {a^6 \bar V}
\end{pmatrix} 
\ee
a metric in the internal field space in which the time derivatives of
the scalar fields  are the coordinates. Summation over the repeated 
indexes $(I,J,\ldots)$ is implicitly assumed. The evolution of the
homogeneous system becomes therefore equivalent to that of a nonlinear
sigma model with biscalar interactions encoded in a curved field
manifold.

In spite of the appearances, the action~(\ref{Smetric})
\textit{completely captures} the dynamics of the original background
field $\bar\phi$ in the potential $\bar V$. This becomes clear by
noticing that the $G_{\chi\chi}$ component of the field-space metric is
inversely proportional to $a^6 \bar V$, where the presence of the scale
factor---which is just a manifestation of the vectorial origin of
this field---in this expression is extremely important.\footnote{
Indeed, if the (non-canonical) kinetic term for the $\chi$ field did not
involve this contribution, the corresponding equation of motion would read 
\be
\p_t\l(\f{a^3\dot{\bar\chi}}{\bar V}\r) = 0 \,,
\ee
leading to an equation of motion for $\bar\phi$, 
\be\label{eq:wrong}
\ddot{\bar\phi}+3 H\dot{\bar\phi} = -\f{ \bar V'}{a^6} \,,
\ee
that obviously differs from the usual the Klein-Gordon equation in an
FLRW spacetime. Actually, for a (quasi-) de Sitter spacetime, the
right-hand side of the last expression vanishes exponentially fast as
the Universe expands, leading effectively to the freezing of the
background scalar field.}

\subsection{Covariant first-order perturbations}

Even though the full nonlinear
theories~(\ref{eq:minimal_coupling_scalar_act})
and~(\ref{eq:minimal_coupling_equiv_action}) are completely
equivalent~\cite{Finn:2020lyw}, it is instructive to demonstrate how
this comes about forr first-order fluctuations on top of an FLRW
background. To this end, let us consider linear perturbations over the
fixed metric $\bar g_{\mu \nu}$, 
\be
g_{\m\n}\simeq \bar g_{\m\n}+\d g_{\m\n}\,, 
\ee
and decompose the scalar and vector fields  into a background quantity
and a small perturbation over it,
\be
\phi \simeq  \bar\phi +\d \phi \,, \hspace{5mm} B_\m \simeq \bar B_\m +
\d B_\m \ .
\ee 
Plugging this decomposition into the equations~(\ref{eq:phifield_full})
and~(\ref{eq:constraintB}) for the scalar and vector fields, we obtain
the background equations of motion at zeroth order in the perturbative
expansion, 
\be
\bar \nabla_\m \bar B^\m =\f{\sqrt 2 }{M^2} \bar V \,,\hspace{10mm} \bar
\Box \bar \phi = \bar F \bar V'  \ ,   
\ee
as well as those of the fluctuations at first order,
\begin{align}
&\bar \nabla_\m {\cal B}^\mu = \f{\sqrt 2 }{M^2}\left(\bar V'\d\phi+\f 1
2 \bar V \bar g^{\m\n}\d g_{\m\n}\right) \ , \\
&\bar\nabla_\m \,\delta(g^{\mu\nu} \partial_\nu \phi)
=\bar F^2\left(\bar V''\d \phi + \frac12 \bar V'\bar g^{\m\n}\d
g_{\m\n}\right) +  \bar V' \delta ( F^2) \ ,
\end{align}
with 
\begin{eqnarray}
&&  \bar F=F(\bar B,\bar \phi)\ , \\ &&{\cal B}^\mu=
\delta(g^{\mu\nu}  B_\nu ) =\bar g^{\m\n} \d B_\n -\f 1
2M^{\mu\lambda\nu\sigma}\bar B_\n \d g_{\lambda\sigma}\,, \\
&&\delta(g^{\mu\nu} \partial_\nu \phi)=\bar g^{\m\n}\p_\n\d\phi -\f 1 2
M^{\mu\lambda\nu\sigma}\p_\n\bar\phi\d g_{\lambda\sigma}\,, \\
&&M^{\mu\lambda\nu\sigma}=\bar g^{\m\lambda}\bar g^{\n\sigma}+\bar
g^{\m\sigma}\bar g^{\n\lambda}-\bar g^{\m\n}\bar g^{\lambda\sigma}\,,\\
&& \delta (F^2) = - \frac{M^4}{\bar V^2}  \biggl (\nabla_\lambda \bar
B^\lambda \nabla_\mu \bar B_\nu \delta g^{\mu \nu} \nonumber \\
&&\qquad\qquad + \nabla_\lambda \bar B^\lambda \bar B_\mu  \nabla_\nu
\delta g^{\mu \nu}  - \nabla_\lambda \bar B^\lambda \nabla_\sigma \delta
B^\sigma   \\ && \qquad\qquad- \frac{1}{2} \bar B_\nu \nabla_\lambda
\bar B^\lambda \nabla^\nu \delta g_{\,\sigma}^\sigma + \frac{
(\nabla_\lambda \bar B^\lambda)^2 \bar V'}{\bar V} \delta \phi\biggr
)\nonumber \ .
\end{eqnarray}
The structure of the scalar field equations in this set of expressions
is highly suggestive. Upon employing the equation of motion for $\bar
B_\m$ and $\d B_\m$, we obtain $\delta (F^2)=0$. Further, setting $\bar
F=1$, the above boils down to the usual background and first-order
perturbation equations for the original inflaton field $\phi$, namely
\begin{align}
&\bar \Box \bar \phi =\bar V' \ , \\
&\bar\nabla_\m \delta(\bar g^{\mu\nu} \partial_\nu \phi) = \bar V'' \d
\phi+\frac12 \bar V'\bar g^{\m\n}\d g_{\m\n} \,.
\end{align}
The same applies to the Einstein equations at the corresponding orders,
namely
\begin{align}
& M_P^2\bar G_{\m\n} = \bar T^{(\phi)}_{\m\n} +\bar  T^{(B)}_{\m\n}
\ ,\\ 
& M_P^2\,\d G_{\m\n} =\delta T^{(\phi)}_{\m\n} + \delta T^{(B)}_{\m\n} \
,
\end{align}
with 
\begin{eqnarray} 
&&\bar T^{(\phi)}_{\m\n} = \p_\m \bar\phi \p_\n \bar\phi -\f 1 2 \bar
g_{\m\n} \p^\sigma  \bar\phi\p_\sigma\bar\phi \ , \label{eq:Tscalar}\\
&&\bar T^{(B)}_{\mu\nu} =\sqrt{2}M^2 \left(\bar B_{(\m}\p_{\n)}- \bar
g_{\mu \nu} \bar B^\lambda \p_\lambda \right)  \bar F- \bar g_{\mu \nu}
\bar F^2\, \bar V,\label{eq:Tvector}
\end{eqnarray}
and
\begin{eqnarray} 
\d T_{\m\n}^{(\phi)}&=&2\p_{(\m} \bar\phi \p_{\n)} \d\phi-\bar
g_{\m\n}\bar g^{\rho\sigma}\partial_{(\rho}\bar\phi \partial_{\sigma)}\d
\phi \nonumber\\ &+&\f 1 2
\left(\bar g_{\m\n}\bar g^{\rho\kappa}\bar g^{\sigma\lambda}-\bar
g^{\rho\sigma}\d^{\kappa}_{\m}\d^{\lambda}_{\n}\right)\partial_{\rho}\bar\phi
\partial_{\sigma}\bar\phi \d g_{\kappa\lambda} \ , \\
\label{eq:Tvector_pert}
\delta T^{(B)}_{\mu\nu} &=&\sqrt{2}M^2 \big[  \delta\left(
B_{(\m}\p_{\n)}- g_{\mu \nu}  B^\lambda \p_\lambda \right) \bar F 
\nonumber  \\  &+& \left(\bar B_{(\m}\p_{\n)}- \bar g_{\mu \nu} \bar
B^\lambda \p_\lambda \right)  \delta  F \big]  \nonumber \\ &-&\, \bar
g_{\mu \nu}  \bar V \delta (F^2)\, - \bar F^2\, \delta (g_{\mu \nu}
V)\ .
\end{eqnarray}
Again, using the equations of motion for $\bar B_\m$, $\d B_\m$ and $\d
\phi$ and setting $\bar F=1$, the energy-momentum
pieces~(\ref{eq:Tvector}) and~(\ref{eq:Tvector_pert}) collapse to the
standard contributions from the scalar potential $V$
\begin{eqnarray}
&& \bar{T}_{\mu \nu}^{(B)} = - \bar g_{\m\n} \bar V \, ,\\
&& \delta T_{\mu \nu}^{(B)}  = - \bar{V} \delta g_{\m\n} - \frac{1}{2}
\bar g_{\m\n}  \bar{V}\delta g^\lambda_{\,\lambda} -\bar  g_{\m\n}
 \bar{V}'\delta \phi\ ,
\end{eqnarray}
giving rise, therefore, to the background and perturbed Einstein
equations in agreement with those in the original
theory~(\ref{eq:minimal_coupling_scalar_act}).

\section{The (Geometry of) Inflationary observables}
\label{sec:observables}

Having demonstrated the equivalence of the two theories both at
background level and first order in perturbation theory, we proceed now
to determine the inflationary observables in the uplifted theory. To do
so, let us consider the scalar curvature associated with the
two-dimensional field-space metric, namely 
\be
\mathcal R=G^{I J} \left( \partial_K \Gamma^K_{I J} - \partial_J
\Gamma^L_{L I} + \Gamma^K_{K N} \Gamma^N_{I J} - \Gamma^K_{J N}
\Gamma^N_{K I} \right)\ , 
\ee
with
\be
\Gamma^K_{I J}  = \frac{1}{2} G^{K L} (\partial_I G_{L J} + \partial_J
G_{L I} - \partial_L G_{I J}) \ . 
\ee
A straightforward computation reveals that 
\be
\label{eq:R_pot_rel_1}
\mathcal R = \f{\bar V''}{\bar V} - \f 3 2 \l(\f{\bar V'}{\bar V}\r)^2 \
.
\ee
As expected the resulting expression comprises only terms with
derivatives of the potential, without any explicit dependence on the
scale factor or the parameter $M$. We can go a step further though, by
noticing that the same holds true for the  following quantity, which, in
a clear abuse of language, we will coin the ``mean
curvature''\,\footnote{Strictly speaking, the mean curvature for the
case at hand should be defined as 
\be
\langle \mathcal R\rangle = \f{\int d\varphi  \sqrt{\det
G(\varphi)}\mathcal R(\varphi)}{\int d\varphi\sqrt{\det G(\varphi)}} \ ,
\ee
whereas the quantity $\mathscr R$ we introduced in the main text is only
normalized to the square root of the field-space metric determinant
rather than to the volume. }
\be
\label{eq:R_pot_rel_2}
\mathscr R =\f{1}{\sqrt{\det G(\bar\phi)}} \limitint^{\bar\phi} d\vp\,
\sqrt{\det G(\vp)}\, \mathcal R(\vp)= \f{\bar V'}{\bar V}  \ .
\ee

When the formulas~(\ref{eq:R_pot_rel_1}) and~(\ref{eq:R_pot_rel_2}) are
written in terms of the first and second slow--roll parameters 
\be
\ep = \f{M_P^2}{2} \l(\f{\bar V'}{\bar V}\r)^2 \,\hspace{7mm}\eta =
M_P^2\f{\bar V''}{\bar V} \ ,
\ee
they allow to fully ``geometrize'' the usual expressions for the
amplitude of the primordial spectrum of density fluctuations,
\be
\f{\bar V}{\l(M_P\mathscr R\r)^2} =  2.5\times 10^{-7} M_P^4 \ ,
\ee
and the associated spectral tilt,
\be\label{ns}
n_s = 1+2 M_P^2\mathcal R \ .
\ee
A completely flat Harrison--Zeldovich spectrum corresponds therefore to
a flat field manifold, being the deviation from it proportional to the
scalar curvature ${\cal R}$, which is generically restricted to be
negative at horizon exit in order to satisfy the observational
constraint $n_s < 1$ \cite{Planck:2018jri}. This indeed is not the end
of the story, since the tensor--to--scalar ratio is  also related to the
field-space as
\be \label{r}
r  =  8 \l(M_P\mathscr R\r) ^2 \ .  
\ee 
In this geometrized picture, the evolution equations for the first two
slow-roll parameters \cite{Hoffman:2000ue} can be replaced by those of
the scalar $\mathcal R$ and mean $\mathscr R$ curvatures. Assuming
slow-roll (see Eq.~(\ref{eq:slow-roll_curv}) below), these take the
form\,\footnote{The first equation can be shown to be equivalent to
\be 
\label{flowmathscrR}
\frac{d  {\mathscr R}'}{dN}  =\frac{d }{dN}\left({\cal R} + \frac12
{\mathscr R}^2\right)\ .
\ee}
\be 
\label{eq:flow_explicit}
\frac{d {\cal R}}{dN}= - M_P^2  {\mathscr R}^2   \left({\cal R} +
\frac12 {\mathscr R}^2   \right) +   M_P^2  {\mathscr R} {\mathscr R}''\
,  
\ee
and 
\be 
\label{flowR}
\frac{d \mathscr R}{d N}=M_P^2 \mathscr R \left(\mathcal R + \frac12
{\mathscr R}^2\right) \ ,
\ee
with 
\be
N=\int \frac{d\phi}{M_P^2 {\mathscr R}} 
\ee
the number of inflationary $e$-folds, and 
\be
\label{eq:slow-roll_curv}
 M_P\, \mathscr R\ll \sqrt{2}\,.
\ee
In non-geometrical terms, the latest expression corresponds to a
``velocity hierarchy'' in field space, with the fictitious
field-direction evolving faster than the real one, \ie $\dot
{\bar\phi}^2 \ll M^2\dot{\bar\chi}/\sqrt{2}a^3$,\footnote{ This is just
the standard requirement $\dot{\bar \phi}^2\ll \bar V$ written in the
Eisenhart language, i.e. by expressing the potential in terms of
$\bar\nabla\cdot \bar B$ through~Eq.~(\ref{eq:constraintB}) and then
trading $\bar B_0$ for $\bar \chi$ using Eq.~(\ref{eq:def_chi}). }. This
reduces to the standard slow-roll condition $\epsilon<1$ upon using the
background equation of motion for  $\chi$.  

{\renewcommand{\arraystretch}{2}
\setlength{\tabcolsep}{10pt}
\begin{table*}[t]
  \centering
\begin{tabular}{cccc l c}
\hline
$a_0$ & $a_1$ & $a_2$& &\hspace{10mm}  Potential & Model \\
\hline \hline
$\frac{2M_P^2}{f^2} $ & $0$ & $-1/2$ && $V(\phi)= V_0\, \sech^2
\,\frac{\phi}{f}$ & Constant ${\cal R}$ inflation \\
$ -\frac{M_P^2}{f^2}  $ & $ 0 $ & $1 $ & &  $V(\phi)= V_0\,
\sinh\frac{\phi}{f}$ 
& Hyperbolic inflation \cite{Basilakos:2015sza} \\
$ \frac{M_P^2}{2f^2} $ & $ 0 $ & $1/2 $ & &  $V(\phi)=
V_0\,\left(1-\cos\frac{\phi}{f}\right)$ & Natural inflation
\cite{Freese:1990rb} \\
\hline
\end{tabular}
\caption{Equivalent potentials shapes for different choices of the
Taylor coefficients $a_0,a_1,a_2,\ldots$ in the expansion
\eqref{flowRapprox}. }
\label{tab:1}
\end{table*}}

\section{Lifts, Flows \& Equivalence Classes}
\label{sec:lifts}

The structure of Eqs.~(\ref{flowmathscrR})-(\ref{flowR}) invites us to
formulate inflation as a theory of fixed points in curvature space (for
simular approaches see for instance
Refs.~\cite{Hoffman:2000ue,Kinney:2002qn,Boubekeur:2014xva}). Indeed, as
we will see in what follows, an inflationary period corresponds to the
slow-roll of the field towards or away from a fixed point. By simple
inspection, we recognize two fixed points. The first one at 
\be
\mathscr  R=\textrm{constant} \,, \hspace{10mm}\mathcal R+\frac12
{\mathscr R}^2=0\,,
\ee
leads to a power-law expansion with 
\be
n_s=1-\frac{r}{8}  \,.  
\ee
In the original single-field representation, this corresponds to an
exponential runaway inflationary potential \cite{Schwarz:2004tz}. Beyond
being currently disfavoured by Planck at more than 2$\sigma$ C.L.
\cite{Planck:2018jri}, this scenario has a graceful exit problem in the
absence of additional exit mechanisms. 

The second fixed point at 
\be
{\mathscr R}=0\,,\hspace{10mm}  \mathcal R=\textrm{constant}\,, 
\ee
leads to a constant spectral tilt and a vanishing tensor-to-scalar ratio,
\be
n_s=\textrm{const.}\,, \hspace{10mm} r=0\,,
\ee
being stable against perturbations for ${\cal R}>0$ (or $n_s>1$) and
unstable for ${\cal R}<0$ (or $n_s<1$). In the single field language,
this corresponds to an inflationary setting involving a local minimum or
maximum (i.e. an unstable equilibrium point). 

The limiting case  ${\mathscr R}=0$, ${\cal R}=0$ ($n_s=1$, $r=0$) gives
 rise to an exact de Sitter expansion. According to the behaviour of
$\mathscr R$ and $\mathcal R$ in the vicinity of this specific solution,
we can distinguish several equivalence classes:
\begin{itemize}
\item \textit{Exponential fixed-point approach}: In this case, the de
Sitter fixed point is reached exponentially fast as $\phi\to \infty$
\be
\hspace{5mm}{\mathscr R}\sim \frac{2\alpha }{M_P} e^{-\frac{\alpha
\phi}{ M_P}} \ , \hspace{5mm}\mathcal R+\frac12 {\mathscr R}^2\sim 
-\frac{\alpha}{M_P} {\mathscr R} \ .
\ee
This class includes the Starobinsky model of
inflation~\cite{Starobinsky:1992ts}, Higgs
inflation~\cite{Bezrukov:2007ep} in different gravity
settings~\cite{Rubio:2018ogq}, and the $\alpha$-attractor
scenarios~\cite{Kallosh:2013hoa,Galante:2014ifa}. In the large $N$
limit, the spectral tilt and tensor-to-scalar ratio boil down to the
well known expressions
\be
n_s\simeq 1-\frac{2}{N}\,,
\hspace{10mm} r \simeq  \frac{8}{\alpha^2N^2} \ .
\ee
\item  \textit{Rational fixed-point approach}: In this equivalence
class, the de Sitter fixed point is approached as
\be
 {\mathscr R}=\frac{\beta}{\phi-\phi_0} \ , \hspace{8mm} \mathcal
 R+\frac12 {\mathscr R}^2= -\frac{1}{\beta}{\mathscr R}^2\ , 
\ee
either at a finite value $\phi_0\neq 0$ or at $\phi\to\infty$
($\phi_0=0$), with $\beta$ an ${\cal O}(1)$ dimensionless parameter. For
$\beta>0$, this  type of scaling accommodates all chaotic inflation
scenarios. For $\beta<0$, it describes inverse monomial potentials, like
those appearing for instance in quintessential \cite{Ratra:1987rm} or
tachyon inflation settings \cite{Feinstein:2002aj}. In the large $N$
limit, the associated spectral tilt and tensor-to-scalar ratio become
\be
n_s\simeq 1-\frac{\beta+2 }{2N}\,,\hspace{10mm} r \simeq  \frac{4|\beta|
}{N}\ .
\ee
The specific case $\beta=-2$ leads to a fully scale-invariant spectrum 
with $n_s=1$. 
\item  \textit{Monomial fixed-point approach}: In this equivalence
class, the de Sitter fixed point is approached as
\be
\hspace{7mm} {\mathscr
R}=\frac{\beta}{M_P}\left(\frac{\phi}{M_P}\right)^n \ , \hspace{2mm}
{\mathcal R}+\frac12 {\mathscr R}^2\sim \frac{n \beta}{M_P^2}
\left(\frac{M_P{\mathscr R}}{\beta} \right)^{\frac{n-1}{n}}\ .
\ee
For $n<0$, the situation is similar to the previous cases, with the
fixed point at $\phi\to \infty$ approached now as an inverse power law.
On the other hand, for $n>1$, the fixed point is approached at small
field values.  In the large $N$ limit, the spectral tilt and
tensor-to-scalar ratio become
\be
\hspace{5mm} n_s \simeq 1+\frac{\alpha}{N}-\frac{r}{8}\ ,\hspace{5mm} r
\simeq 8 \beta ^2 \left(\frac{\alpha}{2n \beta N}\right)^{\alpha}\ ,
\ee
with $\alpha=2 n/(n-1)$.
\item \textit{Parametric fixed-point approach:} The de Sitter limit can
be also approximately realized if the quantity $M_P^2\left(\mathcal R +
\frac12 {\mathscr R}^2\right)$ remains small and slowly varying in a
given field range. To illustrate this, let us consider a relation 
$M_P^2\left(\mathcal R + \frac12 {\mathscr R}^2\right)=-f(M_P {\mathscr
R})$ with $f$ an arbitrary function and $M_P {\mathscr R}\ll1$, as
suggested by the experimental constraints on the tensor-to-scalar ratio.
Expanding the right hand side of this expression in power series, we
have
\be\label{flowRapprox}
f(M_P {\mathscr R})=a_0 +a_1 M_P {\mathscr R}
+a_2 M^2_P {\mathscr R}^2 +\ldots \hspace{-4mm}
\ee
This parametrization covers a plethora of inflationary models with
completely different predictions. In particular, given specific choices
of the Taylor coefficients $a_0,a_1,a_2,\ldots$ we can easily
reconstruct seminal examples in the literature such as natural or
hyperbolic inflation, cf. Table \ref{tab:1}. Note also that, in order to
support a sufficiently long period of inflation, the $a_0$ parameter
must be much smaller than one. This typically requires the appearance of
a new scale $f\gg M_P$ that can be identified with the curvature radius
of the field space manifold. 
\end{itemize}

\section{Conclusions}
\label{sec:conclusion}

In this paper we have presented a novel method to study inflationary
models using the Eisenhart lift formalism for field theories.  In this
context, we have demonstrated that the main inflationary observables can
be expressed in terms of  geometrical objects describing the
two-dimensional field-derivative manifold  of the uplifted theory, for
which the time derivatives of the scalar and temporal component of the
fictitious vector play the role of the coordinates. Furthermore, this
method has the potential to be extended to multi-field inflationary
settings, allowing for an even wider range of models with completely
different predictions to be captured by the fixed point picture. The
examples presented in this paper were previously studied in the
literature, and we anticipate that this approach will lead to further
insights and advancements in our understanding of inflation and the
early Universe. Overall, our results demonstrate the power and
versatility of the Eisenhart lift method for field theories in the
context of inflationary cosmology.

\section*{Acknowledgements}

We are grateful to Mikhail Shaposhnikov and especially Sebastian Zell
for comments on the manuscript. The work of M.~M. was supported by a
Minerva Fellowship of the Minerva Stiftung Gesellschaft für die
Forschung mbH and in part by the Israel Science Foundation (grant No.
741/20) and by the German Research Foundation through a German-Israeli
Project Cooperation (DIP) grant ``Holography and the Swampland.''  J.~R.
is supported by a Ram\'on y Cajal contract of the Spanish Ministry of
Science and Innovation with Ref.~RYC2020-028870-I. 

\bibliographystyle{utphys}
\bibliography{References}

\end{document}